\documentclass[12pt]{iopart}

\usepackage{graphicx}
\usepackage{iopams}
\usepackage{hyperref}
\usepackage{upgreek}
\usepackage[caption=false]{subfig}

\newcommand{\rmt}[1]{\textrm{\scriptsize{#1}}}		
\newcommand{\eqref}[1]{equation~\eref{#1}}	

\begin{document}
\title[Studies of atomic diffusion in Ni-Pt solid solution by XPCS]{Studies of atomic diffusion in Ni-Pt solid solution by X-ray Photon Correlation Spectroscopy}

\author{Markus Stana$^1$, Michael Leitner$^{1,2}$, Manuel Ross$^1$ and Bogdan Sepiol$^1$}

\address{$^1$ Universit\"at Wien, Fakult\"at f\"ur Physik, Strudlhofgasse 4, 1090 Wien, Austria}
\address{$^2$ FRM-II, Technische Universit\"at M\"unchen, Lichtenbergstra{\ss}e 1, 85747 Garching, Germany}
\ead{markus.stana@univie.ac.at}

\begin{abstract}
Atomic scale X-ray Photon Correlation Spectroscopy (aXPCS) was used to study atomic diffusion in the Ni$_{97}$Pt$_{3}$ solid solution with both a single crystal and a polycrystalline sample. Different jump diffusion models are discussed using experimental results and Monte Carlo simulations. 
The sensitivity of aXPCS experiments to short-range order (in this case governed by a strong Pt-Pt repulsive force) is demonstrated. 
The activation energy of 2.93(10)\-eV as well as diffusivities in the range of $10^{-23}\-$m$^2\-$s$^{-1}$ at 830~K agree very well with the results of tracer diffusion studies at much higher temperatures.
\end{abstract}

\submitto{\JPCM}
\maketitle

\section{Introduction}
\label{sec:intro}

A variety of well established methods can be applied to investigate diffusion in solid state systems \cite{mehrer2007diffusion}. However, studies of diffusion mechanisms on the atomic scale are still challenging. Atomic scale X-ray Photon Correlation Spectroscopy (aXPCS) can provide new insight into this field. 

XPCS is the X-ray analogue of the very successful technique of dynamic light scattering or Photon Correlation Spectroscopy \cite{brown1993dynamic}. PCS is well established to characterize the equilibrium dynamics of soft condensed matter by determining the intensity autocorrelation function of the scattered laser light as a function of delay time and wave-vector transfer. Similarly, XPCS \cite{sutton2008}, using coherent synchrotron radiation instead of visible laser light, is being widely used for studies of soft matter dynamics where objects are in the nanometre range (see e.g.~\cite{gruebel2008, lu2010temperature, cipelletti2011, chushkin2012}). Attention has also been directed towards studies of metal systems as shown by the feasibility test on the (100) Bragg peak of Cu$_{3}$Au \cite{sutton1991} or of critical dynamics in Fe$_{3}$Al \cite{brauersutton1995}. Because of the short X-ray wavelength, XPCS seems to be predestined for studies at atomic length scales. These types of experiments have, however, never been reported in the literature until a successful demonstration of the capability of XPCS to resolve single atomic motion in condensed matter \cite{leitner2009atomic}. Particularly, we have clarified the mechanism of atomic dynamics in Cu$_{90}$Au$_{10}$ \cite{leitner2009atomic, leitner2011quasi} and studied dynamics in a metallic glass \cite{PhysRevB.86.064202}. 

The goal of this paper is to describe in detail the atomic motion of dilute platinum atoms in a nickel matrix at relatively low temperatures. 
The Ni$_{97}$Pt$_{3}$ intermetallic alloy is a perfect candidate for studying diffusion of small quantities of dilute atoms due to its outstanding intensity of diffuse scattering even for low solute concentration levels. 
Moreover, some profound questions can be addressed regarding interactions between solute atoms and interactions of solute atoms and vacancies. It is well known that vacancies are vehicles of diffusion in alloys \cite{mehrer2007diffusion}, but the interactions between vacancy and solute atom can radically influence the motion of the latter. Our objective is to analyse these interactions with the help of aXPCS. 
These questions were addressed in the past by a number of methods like M{\"o}ssbauer spectroscopy, nuclear resonant scattering or quasi-elastic neutron scattering \cite{voglsepiol2005}. The elementary diffusion steps were investigated for self diffusion \cite{voglpetry1989}, for impurity diffusion with consideration to the interaction with vacancies \cite{steinmetz1986}, for hydrogen in intermetallic hydrides \cite{hempelmann2009} and in intermetallic B2 \cite{voglsepiol1994, feldwisch1995, kaisermayr2000} and D0$_3$ \cite{sepiolmeyer1998} phases. All methods mentioned above have in common that they are limited to only few selected isotopes and have a poor energy resolution which forces measurements at very high temperatures (apart from hydrogen). The time-domain technique of aXPCS is, however, free from these restrictions.    

The important aspect of this paper is studying diffusion in single crystal and polycrystalline samples of the same composition. 
For methods studying spatially resolved atomic motion like aXPCS, single crystal samples provide the possibility to investigate different crystallographic directions and therefore more information. It is, however, sometimes difficult or even impossible to obtain single crystals of satisfactory quality. Showing that studies on polycrystals can, in whole or in part, provide information on a comparable level when using a coherent method, a new field of materials could be investigated with aXPCS. Ni$_{97}$Pt$_{3}$ seems to be a predestined object for such a study.

\section{Theory}
\label{sec:theory}

Atomic structures can be investigated at their elemental scale. Still, monitoring a single atom in a bulk sample on time scales of atomic diffusion it is not possible in real space. It is however possible to measure dynamics in reciprocal space which corresponds to these dynamics in real space. 

Usually, reconstruction of real space from a scattering image is very challenging due to the phase problem. Fortunately, one can ignore the details of the exact structure factor, but use the time fluctuations of the intensity pattern to study dynamics of the material with XPCS. This is possible because the intensity autocorrelation function $g^{(2)}(\vec{{q}},t)$ and the autocorrelation function of the electric field $g^{(1)}(\vec{{q}},t)$ are connected by the Siegert relation for Gaussian scattering processes \cite{lemieux1999investigating}. Therefore it is sufficient to measure the intensity autocorrelation function to gain information about the dynamics in real space 
\begin{equation}
 g^{(2)}(\vec{{q}},\Delta {t}) = \frac{\langle I(\vec{{q}},{t}) \cdot I(\vec{{q}},{t} + \Delta {t}) \rangle}{\langle I(\vec{{q}},{t}) \rangle^2},
\label{eq:int_corr}
\end{equation}
where $I(\vec{{q}},{t})$ is the observed intensity at wave-vector $\vec{{q}}$ and time $t$ and the brackets $\langle ...\rangle$ denote an average over the ensemble. In XPCS and in many other types of experiments, the time average of the investigated quantity is calculated from the measured data. 

XPCS is the time Fourier-transformed counterpart of coherent quasi-elastic neutron scattering (QENS). It can be shown that the correlation time $\tau(\vec{{q}})$ corresponds to the energy line broadening \cite{hempelmann}  $\Gamma(\vec{{q}})$ via $\tau(\vec{{q}}) \propto  \Gamma(\vec{{q}})^{-1}$. Therefore concepts developed for QENS can be adopted to XPCS. 

For a coherent method like aXPCS two atoms of the same species exchanging their positions do not affect in any way the scattering pattern, as the configuration of the coherently illuminated volume before and after the jump is exactly the same. Compared to non-coherent methods it is, however, necessary to account for the short-range order, i.e.~for the influence of the surroundings on the movement of each distinct atom. This phenomenon has first been described qualitatively for coherent neutron scattering in liquids by de Gennes \cite{de1959liquid}. The physical principle for the so-called de Gennes narrowing is based on the fact that peaks in the structure factor of liquids occur at wave-vectors corresponding to the most probable interatomic separation, i.e.~they are responsible for the highly correlated and long living atomic arrangements. The same argument holds for the short-range order intensity ($I_{\rmt{SRO}}$) in alloys where the higher diffuse intensity between Bragg peaks corresponds to a longer lifetime of certain configurations of solute atoms. 

The simplest case of Lorentzian line-shapes with energy-domain methods \cite{voglsepiol2005} corresponds to an exponential decay of the intensity autocorrelation function in time
\begin{equation}
 g^{(2)}(\vec{{q}},\Delta {t}) = 1+ \beta \exp({-2 \Delta {t} / \tau (\vec{{q}})}).
\label{eq:g2}
\end{equation}
The intensity autocorrelation function for perfectly coherent radiation would decline from a value of 2 to 1. However, because of an only partially coherent X-ray beam in the experimental setup the amplitude $\beta$ is diminished. 
The correlation time $\tau(\vec{{q}})$ is given by 
\begin{equation}
 \tau (\vec{{q}}) = \tau_{0} \cdot \frac{I_{\rmt{SRO}}(\vec{{q}})}{\sum_{n} P_{{n}} ~\sum_{\Delta \vec{a}_{{nj}}} \left(1-\exp\left({\rmi\,\vec{{q}} \cdot \Delta \vec{a}_{nj} }\right)\right)}.
\label{eq:tau}
\end{equation}
This follows from the theory describing the short-range order and coherent dynamics in a linear approximation as derived by Sinha and Ross \cite{Sinha198851}. It is discussed in more detail in Leitner and Vogl \cite{leitner2011quasi}. $\tau_0$ is the average time between jumps of an atom and therefore corresponds to the inverse of the average jump frequency $\nu_0^{-1}$. The probability for an atom to jump to one of the lattice sites of a certain neighbouring shell is given by $P_{{n}}$, with $n$ being the number of the appropriate coordinate shell (1 for nearest-neighbour, 2 for next-nearest-neighbour and so on). The vector $\vec{a}_{{nj}}$ points from the origin to the $j$th site in the $n$th shell. It can be easily seen that the inverse correlation time is the sum over the inverse correlation times for each shell corresponding to the particular jump mechanism. Therefore it can also be written as $\tau (\vec{{q}})^{-1} = \sum_{n} \tau_{n} (\vec{{q}})^{-1}$.

In a polycrystalline sample one has to average over all orientations of $\vec{q}$ with respect to the crystal lattice 
\begin{equation}
 \overline{g}^{(2)}({q},\Delta {t}) = 1+ \frac{\beta}{4 \pi} \int_{|\vec{q}|=q}{\rm d}\vec{q} \cdot \exp\left({-2 \Delta {t} / \tau (\vec{{q}})}\right).
\label{eq:g2_poly1}
\end{equation}

It is essential for the method at hand to be able to fit the data set with a function depending on $\Delta t$. As there is no analytical function which satisfies \eqref{eq:g2_poly1}, an approximation has to be made. A similar approach was introduced by Chudley and Elliott \cite{chudell} for liquids. Exchanging integration and exponentiation, an average correlation time $\overline{\tau}({q})$ can be defined as $\overline{\tau} ({q})^{-1} \mathrel{\mathop:}= (4 \pi)^{-1} \int_{|\vec{q}|=q}{\rm d}\vec{q} \cdot \tau (\vec{{q}})^{-1}$. For this expression an analytical solution can be found ($I_{\rmt{SRO}}(\vec{q})$ was also integrated over $|\vec{q}|=q$):
\begin{equation}
 \overline{\tau}({q}) = \tau_{0} \frac{I_{\rmt{SRO}}({q})}{\sum_{n} P_{{n}} ~\sum_{\Delta \vec{a}_{{nj}}} \left(1- \frac{\sin(\vec{{q}} \cdot \Delta \vec{a}_{nj})}{\vec{{q}} \cdot \Delta \vec{a}_{nj}}\right) }. 
\label{eq:tau_chel}
\end{equation}
Using \eqref{eq:tau_chel} the intensity autocorrelation function can be written as: 
\begin{equation}
 \overline{g}^{(2)}({q},\Delta {t}) = 1+ \beta \exp\left({-2 \Delta {t} / \overline{\tau} ({q})}\right).
\label{eq:g2_poly2}
\end{equation}

Numerical calculations show that \eqref{eq:g2_poly2} is a very good approximation for \eqref{eq:g2_poly1} away from Bragg peaks.

\section{Experimental}
\label{sec:Exp}

Nickel and platinum, both of $99.98\%$ purity, were melted by induction heating in a ``cold boat'' in argon atmosphere.  For the preparation of the polycrystalline sample the resulting ingot was cut and rolled between stainless steel plates down to a thickness of 17~$\upmu$m. It was then pre-annealed at a temperature of 773~K for 22~h in a vacuum of about $10^{-7}$~mbar. In order to obtain sufficiently large single crystalline domains the rest of the ingot was heat treated at 1573~K for 140~h resulting in crystallites of about about 5~mm diameter. One domain was oriented, cut with a wire saw and mechanically polished to the final thickness of about 18~$\upmu$m.
The composition of the sample was analysed by energy-dispersive X-ray spectroscopy (EDX) showing 3.0(5)~at.\% platinum content. 

\begin{figure}
\centering
\includegraphics[width=0.5\textwidth]{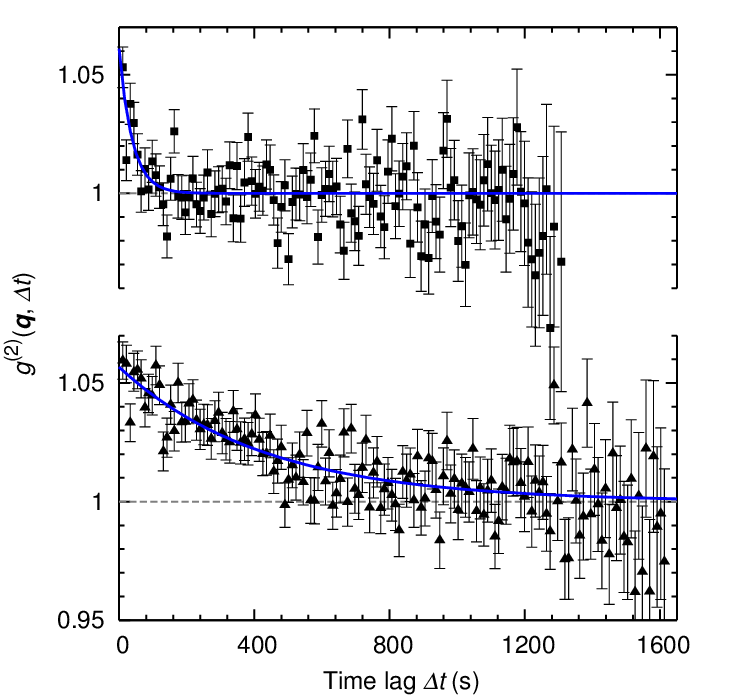}
\caption{\label{fig:g2_exp_fit} Autocorrelation function g$^{(2)}\left(\vec{{q}},\Delta {t}\right)$ in Ni-Pt single crystal orientated with $\left\langle 1 1 0 \right\rangle$ direction parallel to the incoming X-ray beam with azimuthal angle $\phi = 33.2^\circ$ at 855~K (upper plot) and at 815~K (lower plot) fitted with \eqref{eq:g2}.}
\end{figure}
  
The experiment was performed at beamline ID10A of the ESRF using a standard coherent setup with 8~keV photons at a monochromaticity of $\Delta E/E \cong 10^{-4}$. Rollerblade slits were used to select an $8 \times 8\,\upmu$m$^2$ beam section and speckle patterns were recorded with a direct-illumination CCD camera  with $1024 \times 1024$ pixels at a size of $13 \times 13\,\upmu$m$^2$. Prior experience has shown that the CCD camera placed close to the sample (about 0.6~m) is advantageous to maximize the measured signal. It was mounted in a way that allowed it to be moved along a scattering angle $2\Theta$ and an azimuthal angle $\phi$. 
Note that in a scattering experiment the length of the scattering vector is connected to the scattering angle $2\Theta$ via $q=2 |\vec{k}| \sin(2\Theta/2)$, with $\vec{k}$ being the wave vector of the incoming beam. As one can move freely on the Ewald sphere in reciprocal space by changing the detector position in real space, the intensity autocorrelation function was measured for several $\vec{{q}}(2\Theta,\phi)$. In the polycrystalline sample only the scattering angle 2$\Theta$ (i.e.~$\left| \vec{q} \right|$) can influence the correlation time. Therefore measurements at arbitrary azimuthal angles are equivalent. For the single crystal sample an additional second scan for fixed $\left| \vec{q} \right|$ and different azimuthal angles $\phi$ was performed.
Measurements were performed in transmission geometry in a resistively heated vacuum furnace ($p\leq 10^{-6}$~mbar). The temperature was stabilized by a PID temperature controller within a range of 0.1~K. The measurement time at a fixed scattering vector $\vec{q}$ was between 20 and 40 minutes at a frame rate of one frame every 10.9 seconds (exposure plus readout time). Due to the limited intensity of the synchrotron source and low scattering intensity at large scattering angles in the diffuse regime it is beneficial to make the data collected by the CCD camera subject to the so-called droplet algorithm \cite{livet2000using} to detect single photon events. Our implementation of the droplet algorithm is described in \cite{leitner2012data}. These corrected images were evaluated according to \eqref{eq:int_corr}, finding the correlations in the fluctuating speckle intensities (see \fref{fig:g2_exp_fit}) where the autocorrelation function was averaged over all pixels to obtain sufficient statistical accuracy.

All measurements were carried out in the diffuse regime, away from Bragg peaks. Any kind of strain broadening / asymmetry resulting from dislocations, planar faults etc.~are constrained to $q$-regions very close to the Bragg reflections \cite{krivoglaz1996x}. Therefore such effects do not influence the intensities obtained in an aXPCS experiment. The number of vacancies relative to the number of atoms in Ni at 830~K is in the order of $10^{-8}$\cite{koning2003vec}. Therefore the total scattering power of the vacancies and their induced displacements is negligible. An estimation using the Debye model at this temperature  gives a contribution in the order of 10$^{-3}$ originating from thermal diffuse scattering. This contribution can therefore also be neglected. As a result the diffuse intensity is only due to the stochastic occupations of lattice sites by Ni or Pt, and the measured correlation decay therefore corresponds to exchanges of atoms.

Small absolute values of reciprocal vectors correspond to very long correlation times. Instabilities of the experimental setup result in additional intensity variations. This can lead to fault when overlapping with the slowly decaying autocorrelation function caused by the sample itself but is not important if the time-scale of both processes is clearly different. Correlation times measured at very small 2$\Theta$ angles were therefore not used for the evaluation of diffusion models. 

It is vitally important for aXPCS measurements to be performed in thermal equilibrium. To ensure that order relaxation is negligible, multiple measurements were made repeatedly at different detector positions for each sample. Usually the sample relaxes faster when starting from a higher temperature and shows relaxation times of about one hour. After that time the single crystal stayed stable during the whole experiment. The polycrystalline sample, however seemingly relaxed again after about 4~h at 830~K, leading to still longer correlation times. 

Note that the correlation time as measured in an aXPCS experiment quantifies the timescale over which the atomic arrangement changes by a sizeable amount. Even though effects such as enhanced grain boundary diffusion can affect the mass transport over macroscopic distances significantly, the contribution of the grain boundaries to the scattered signal and therefore to the correlation decay is negligible. As a consequence, aXPCS is inherently sensitive only to bulk diffusion. 

To investigate whether there were crystallite growth effects in the polycrystalline sample during the experiment, the sample was studied with electron microscopy. Unfortunately the results were inconclusive. Therefore we assumed that such effects can be neglected as long as no relaxation in the correlation time is detectable.  

\section{Monte Carlo simulations \label{sec:MC}}

To simulate a simple diffusion model, Monte Carlo simulation of diffusion is a valuable approach. Two kinds of atoms were arranged in a proportion of $N_\rmt{Pt}/N_\rmt{Ni} = 3/97$ over an fcc-type lattice with periodic boundary conditions. One vacancy was introduced by randomly emptying a lattice site. 
The simulation consists in iteratively choosing one of the 12 lattice sites of the vacancy's first coordination shell randomly, then, according to the Metropolis algorithm, calculating the probability $p_{i \rightarrow f}= \exp \left(-\Delta E/k_{B}T \right)$, where $\Delta E=E_{f} - E_{i}$ denote the system energies calculated with pair potentials for nearest-neighbours before ($i$) and after ($f$) the jump, respectively, where $k_{B}$ is the Boltzmann constant and $T$ the absolute temperature. A decision to perform a jump or not was taken by comparing $p_{i \rightarrow f}$ with a random number. A unit time (MC step) was defined as a number of trials equal to the number of lattice sites. After equilibrating the system, the actual diffusion simulation was started. After every half Monte Carlo step a discrete Fourier transform of the lattice was carried out and the amplitudes were used to calculate the scattering intensity and the amplitude autocorrelation function $g^{(1)}(\vec{{q}}$, t) for certain $\vec{q}$ values. Finally, using Siegerts relation, the intensity autocorrelation function was calculated and compared with the experimental correlation time. 

The same method was used to calculate short-range order intensities from a reciprocal space map of 64$^3$ voxels with an atom configuration of 4$\times$32$^3$ atoms averaging the intensities over 10000 Fourier transforms.

\section{Results}

In order to distinguish between different diffusion models, measurements of intensity autocorrelation times in polycrystalline and/or in single crystal samples alone are insufficient. As shown in \sref{sec:theory} short-range order intensity plays a crucial role in the interpretation of the measured correlation times. 

\subsection{Short-range order}
\label{sec:ISRO}

The particular importance of local atomic arrangements in alloys for the interpretation of aXPCS measurements follows from \eqref{eq:tau}. Detailed data about SRO like in the case of Cu$_{90}$Au$_{10}$ \cite{schoenfeld1999type} is rarely available. SRO must be actually measured for exactly the same alloy composition and temperature as the sample under study. It is very difficult to deduce short-range order data from experimentally measured intensity data. A detailed description of the diffuse scattering theory can be found in several articles and books (see e.g.~\cite{schwartz1987diffraction, Bernd1999435} and literature cited therein). The factor which actually makes the analysis of diffuse scattering so difficult and time consuming is the separation of SRO scattering (which is of interest to us) and atomic displacement scattering. Simulations showed that in a system with a very high scattering contrast and low solute atom concentration like Ni$_{97}$Pt$_{3}$, the magnitude of those two terms can be of the same order. 
Also terms of higher order in the expansion of diffuse intensity, e.g.~caused by static displacements (Huang scattering) or thermal diffuse scattering, can play a role.

\begin{figure}
\centering
\includegraphics[width=0.5\textwidth]{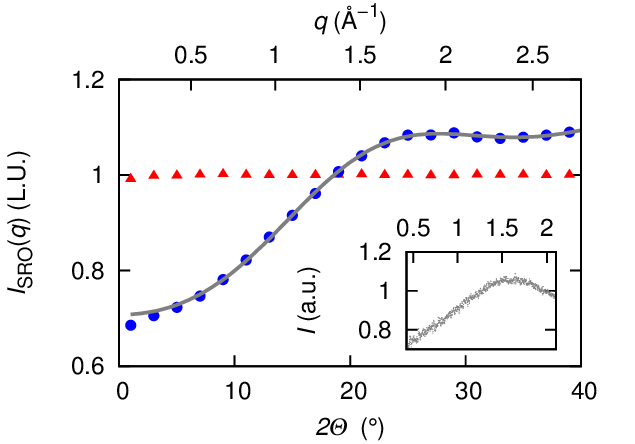}
\caption{\label{fig:ISRO_sc} For a single crystal in $\left\langle 110 \right\rangle$ orientation with azimuthal angle $\phi = 33.2^\circ$ the short-range order intensity was calculated without interaction between the atoms (red triangles) and for strong Pt-Pt repulsion (blue disks). For the latter the short-range order was also calculated from Warren-Cowley parameters gained from the simulation (grey line). Inset shows measured diffuse intensity as a function of the wave-vector transfer.}
\end{figure}

Phase diagrams of the Ni-Pt system \cite{cadeville1986magnetism} show three ordered equilibrium phases similar to Cu-Au: NiPt (L1$_0$ structure), Ni$_{3}$Pt, and NiPt$_{3}$ (both L1$_2$ structures). There was a weak tendency found for a Cu$_{3}$Au-type ordering for Ni$_{1-x}$Pt$_{x}$ with $x=0.2$ composition \cite{PhysRevB.21.5494} which is also indicated by the diffuse intensity for $x=0.03$ as shown in the inset in \fref{fig:ISRO_sc}. There is, however, no data for short-range order configuration on a system with $x=0.03$ available in the literature. 
Note that our main goal remains an atomistic diffusion mechanism and the details of the diffuse scattering intensity are not vitally important. 
Therefore we choose two simple models: (m1) a model with no interaction between the atoms and (m2) a model with strong Pt-Pt repulsing force. The latter was chosen because a L1$_{2}$ ordering type tendency resulting from strong Pt-Pt nearest-neighbour repulsion was assumed to be more likely than from Pt-Pt next-nearest-neighbour attraction or further ranging interactions \footnote{A more thorough investigation of the interaction parameters in the system Ni$_{97}$Pt$_{3}$ based on ab initio calculation will be subject to a following paper.}. The two model systems m1 and m2 were simulated using a Metropolis Monte Carlo algorithm (see \sref{sec:MC}). From the atomic configuration we calculated the Warren-Cowley parameters \cite{PhysRev.120.1648}. In case of 3\% solute atoms with strong repulsive force the calculated parameters were $\alpha_1 = -0.0309$, $\alpha_2 =0.0042$ and $\alpha_3 = 0.0018$. These parameters can be used to calculate the short-range order intensity in Laue units according to
\begin{equation}
 I_{\rmt{SRO}}(\vec{q}) = \sum_{n,j} \alpha_n \cos(\vec{q} \cdot \Delta\vec{a}_{n_j} ),
\label{eq:WaCo}
\end{equation}
where $\Delta\vec{a}_{n_j}$ again is the family of vectors for a certain shell as discussed in \sref{sec:theory}. As shown in \fref{fig:ISRO_sc} the so calculated short-range order intensity is in good agreement with the intensities calculated directly from the Fourier transform of the real-space configuration. 

\subsection{Diffusion mechanism} \label{sec:mech}

The main objective of this paper is to discuss the mechanism which governs diffusion in a Ni-Pt intermetallic alloy. An essential advantage of a microscopic (atomistic) method in diffusion investigation is the possibility to identify such a diffusion mechanism. 
In order to verify which diffusion mechanism operates in a specific intermetallic phase, it is necessary to develop a mathematical model for the diffusion process. This mathematical idea is generated from the knowledge of the system or it is simply a right guess. The mathematical model can be (at least in certain cases) solved analytically and yields the wave-vector dependent correlation time in a single crystal sample (see \eqref{eq:tau}), or the correlation time dependent on the magnitude of the wave-vector in a polycrystalline sample (see \eqref{eq:tau_chel}). A model is determined by the jump probability parameters $P_{n}$. The only free parameter fitted to the experimental data is the inverse average residence time $\tau_0^{-1}$ which corresponds to the jump frequency. 

Needless to say it is not particularly difficult to predict the atomistic mechanism of diffusion in Ni-Pt. It is simply an exchange of a nickel or platinum atom with a vacancy. As the scattering length of platinum considerably exceeds that of nickel and vacancies are very rare in the alloy at the temperatures of measurement, the measurable speckle fluctuations result predominantly from the exchanges of platinum atoms with nickel atoms. This very simple picture enables studies of further subtleties of the diffusion mechanism, such as the influence of interaction between solute atoms or between solute atoms and vacancies.  

A simple and useful tool when discussing a diffusion model involving a vacancy is the so-called encounter approximation \cite{wolf1977}. It has been successfully applied in the interpretation of diffusion in nuclear magnetic resonance and in M{\"o}ssbauer spectroscopy studies\cite{voglsepiol1994,feldwisch1995}. An encounter is here defined as the sum of all exchanges of one and the same atom with one and the same vacancy. As the number of vacancies is much smaller than the number of atoms, the time between one and the same vacancy jumping twice is much smaller than the time between consecutive jumps of one and the same atom. Details during an encounter are thus ``invisible'' for aXPCS and only a net result of one or more exchanges with a vacancy, which can carry an atom to nearest or to further neighbour shells, is accounted. One can assume that encounters are mutually independent and calculate e.g.~diffusivity as a simple random walk of the vacancy. 

In an fcc lattice the first neighbour shell is at a distance of $a/\sqrt{2}$, the second of $a$, the third of  $a \sqrt{3/2}$ and fourth of $a \sqrt{2}$, where $a$ is the lattice constant. The nearest-neighbour shell can be reached by only one exchange between a solute atom and a vacancy. The second, third and fourth shells can, however all be reached after two exchanges with a vacancy. Still farther shells ($n \geq 5$) need three or more exchanges and can therefore be neglected. Also encounters that do not lead to an effective jump are invisible for an aXPCS experiment and are therefore neglected from here on ($P_0$ = 0). We used this to re-normalize the jump probabilities for self diffusion given in \cite{sholl1981} and gained jump probabilities of: $P_1$ = 92.6~\%, $P_2$ = 2.4~\%, $P_3$ = 3.9~\% and  $P_4$ = 0.8~\% (the neglected probability for further jumps sums up to $\sum_n P_{n \geq 5}$ = 0.3~\%). This encounter model (m1), however, describes only systems with negligible interactions between matrix atoms, solvent atoms and vacancies. Therefore we extend the encounter model according to \eqref{eq:tau}, using the SRO intensity for strong Pt-Pt repulsion as discussed in \sref{sec:ISRO}. We use this as a second model for the description of Pt diffusion and call it modified encounter model (m2).

In order to distinguish between the two models, correlation times were measured at a constant temperature of 830~K and at different detector positions corresponding to different $\vec{q}$ and $\left| \vec{q} \right|$ for the single crystal and the polycrystalline sample respectively. 

\begin{figure}[]
  \centering
  \subfloat[Radial scan of single crystal sample]{\label{fig:zwthscan_sc}\includegraphics[width=0.5\textwidth]{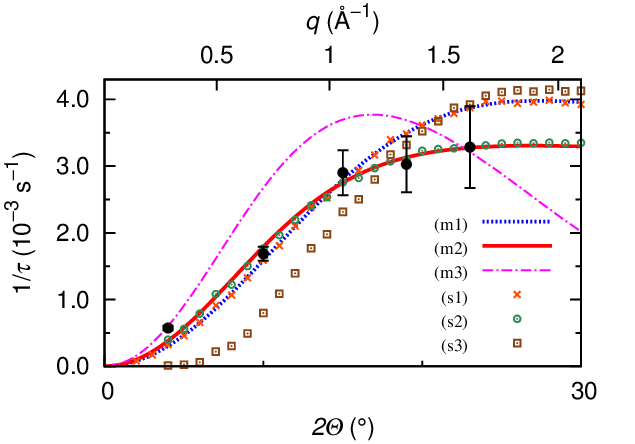}}
  \\ 
  \subfloat[Azimuthal scan of single crystal sample]{\label{fig:phiscan_sc}\includegraphics[width=0.5\textwidth]{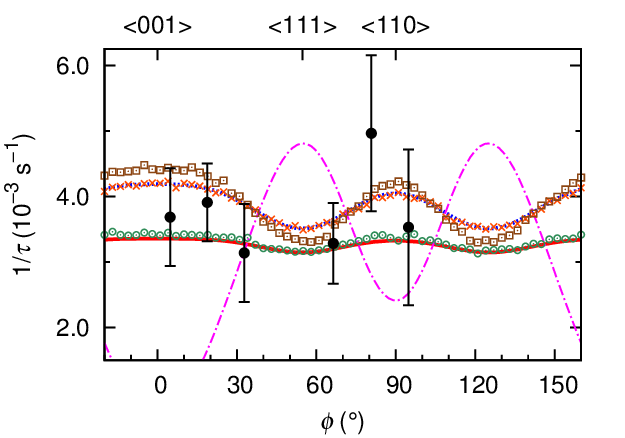}}
  \caption{Reciprocal correlation times in single crystal sample orientated in $\left\langle 110 \right\rangle$ direction at 830~K. The $\left\langle 001 \right\rangle$ direction lies at $\phi$ = $0^{\circ}$. The radial scan (a) was taken at $\phi$ = $33.2^{\circ}$. The azimuthal scan given in (b) was taken at 2$\Theta$ = $23^{\circ}$ . Both datasets were fitted simultaneously with the classical encounter model with no short-range order (m1) and with the modified encounter model with correction for the short-range order calculated for strong Pt-Pt repulsion  (m2).  For comparison, a model with only next-nearest-neighbour jumps (m3) corrected for SRO is also shown. Monte Carlo simulations without interaction potential (s1), with Pt-Pt nearest-neighbour repulsion (s2) and, again for comparison, with Pt-Pt attraction (s3) were carried out.} 
  \label{fig:scans_sc}
\end{figure}

MC simulations sites for different interaction energies between platinum atoms were carried out. One for no interaction between the atoms (s1), one with strong repulsive potential between Pt atoms (s2) and one with an attractive potential between Pt atoms (s3) within the first neighbour shell ($n = 1$). 

The experimentally measured data as well as the data from MC simulations and the models for a diffusion mechanism in the single crystal are shown in \fref{fig:scans_sc}. Note that \emph{the only adjustment parameter} is the jump frequency, which is simply a multiplicative constant. The figure shows that the MC simulation with no interaction potential (s1) complies with the encounter model for a system with no SRO (m1). The model with jump probabilities according to the encounter model and short-range order correction according to \eqref{eq:tau} (m2) gives the same values as the simulated data for strong Pt-Pt repulsion (s2) and describes the experimental data sufficiently well. For this model the  fit produces a jump frequency of $\tau^{-1}_{0}=~3.0(1)\times10^{-3}$s$^{-1}$.  Unfortunately the quality of the measured data was insufficient to draw sound conclusions about solute-vacancy interaction. Obviously a model with only next-nearest-neighbour jumps ($P_2$ = 100~\%, $P_{n \neq 2}$ = 0) (m3) as well as a model with attractive force between Pt atoms (s3) is wrong. 

\begin{figure}[h!]
\centering
\includegraphics[width=0.5\textwidth]{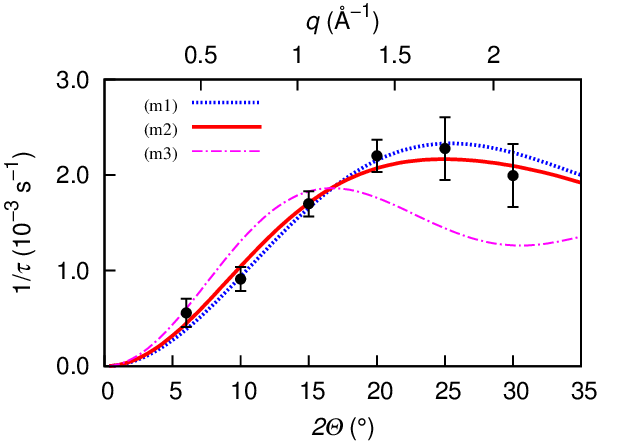}
\caption{\label{fig:chudell} Reciprocal correlation times in polycrystalline sample at 830~K fitted with the classical encounter model (m1), modified encounter model (m2) and for comparison a model with only next-nearest-neighbour jumps (m3).}
\end{figure}

The polycrystalline sample was fitted by \eqref{eq:g2_poly2} and with the same jump probabilities as in the case of the single crystal. The short-range order correction was calculated by numerically integrating \eqref{eq:WaCo} over all orientations of $\vec{q}$. As mentioned in \sref{sec:Exp}, this sample was only quasi-stable for a limited time. It can therefore not be assumed that a sufficient number of small crystallites was covered by the beam during the measurement. For averaging over a small number of crystallites with unknown orientation the Chudley-Elliott model \cite{chudell} is, however, still the best guess. The calculated jump frequency using the modified encounter model as shown in \fref{fig:chudell} was $\tau^{-1}_{0~\rmt{pc}}=~1.92(6)\times10^{-3}$s$^{-1}$. The difference to the single crystal value arises most likely from a difference in sample temperature as a different furnace with a thermocouple at a different position with respect to the sample was used for this measurement. Again a model with only next-nearest neighbour jumps (m3) can obviously not appropriately describe the data measured.

\subsection{Activation energy and diffusion constant}
\label{sec:arrh}

The activation energy $E_{\rmt{A}}$ is the sum of a vacancy formation and migration energy. The relationship between diffusivity, temperature and activation energy can be described fairly accurately by the Arrhenius law \cite{mehrer2007diffusion}. A number of measurements were performed at different sample temperatures. Again measurements for different temperatures where repeated in order to guarantee thermal equilibrium of the sample.

\begin{figure}
\centering
\includegraphics[width=0.5\textwidth]{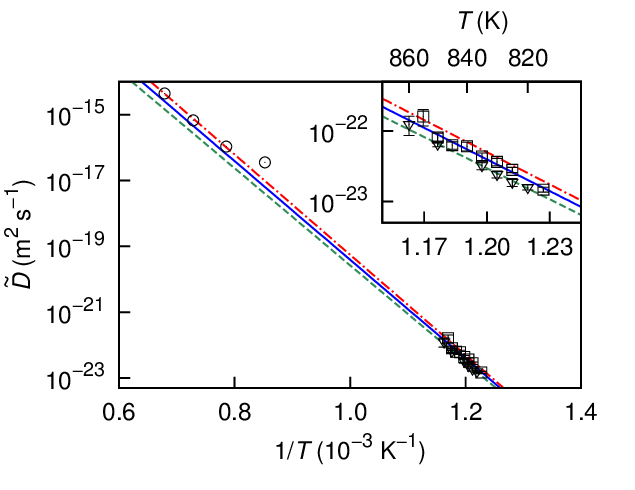}
\caption{\label{fig:arrh} Chemical diffusion coefficients for  polycrystalline (\opensquare) and single crystal sample (\opentriangledown) compared to results from tracer measurements \cite{karunaratne2003interdiffusion} (\opencircle). The green line (dashed) represents the fit for the single crystal sample, the blue line (solid) represents polycrystalline data and the red line (dash-dotted) represents the results of the tracer experiment.}
\end{figure}

Using the Einstein relation \cite{einstein1905bew} and the sum of partial diffusion coefficients into farther shells one can write:
\begin{equation}
 \tilde{D} = \frac{\left < R^2 \right >}{6 \tau_{0}} = \frac{1}{6 \tau_{0}} \sum_{n} P_{n} |\Delta\vec{a}_{n}|^{2},
\label{eq:diffc_einst}
\end{equation}
where $|\Delta\vec{a}_{n}|$ is the radius of neighbour shell $n$. 
For fixed values of $\vec{q}$ one can use \eqref{eq:tau} to calculate the residence time $\tau_0$ from the measured correlation times $\tau(\vec{q})$ for each temperature. The lattice constant for Ni$_{97}$Pt$_{3}$ was calculated using Vegard's law \cite{vegard1921constitution} and corrected for the thermal expansion at 830~K ($a=3.545$~\AA). With this value and using the modified encounter model the chemical diffusion coefficients $\tilde{D}$ can be calculated for each temperature as shown in \fref{fig:arrh}. 

One should notice that a diffusion coefficient measured by the aXPCS technique is, owing to the coherency of the method, the chemical diffusion coefficient $\tilde{D}$. As already emphasized, exchanges between atoms of the same species do not influence the speckle pattern and are not recorded. However, in the case of our particular Ni-Pt solid-solution sample, chemical diffusion and self-diffusion coefficients of platinum are equal in a first approximation. This is simply a result of a low concentration of platinum atoms which makes direct exchanges between them highly improbable.   

Using the Arrhenius law the activation energy as well as the pre-exponential factor $D_0$ can be calculated. Calculated activation energies for the single crystal and the polycrystalline sample, respectively were very similar with $E_{\rmt{A}~\rmt{sc}} = 2.93 (10)$~eV and $E_{\rmt{A}~\rmt{pc}} = 2.97 (18)$~eV. These values are in excellent agreement with literature values for interdiffusion \cite{karunaratne2003interdiffusion}, where the activation energy measured at much higher temperatures (between 900 and 1300$^{\circ}$C) is (291.2 $\pm$ 3.7)~kJ/mol = (3.02 $\pm$ 0.04)~eV. The good agreement between our measurements and the data given in \cite{karunaratne2003interdiffusion} allows us to conclude that there was no grain boundary diffusion contribution in the data given in that reference (apart from the point measured at the lowest temperature). The calculated pre-exponential factors are {$D_{0~\rmt{sc}}$~=~1.5~(+4.6/-1.1)~$\times~10^{-5}\-$m$^2\-$s$^{-1}$} for the single crystal and {$D_{0~\rmt{pc}}$~=~3.7~(+9.2/-3.4)~$\times~10^{-5}\-$m$^2\-$s$^{-1}$} for the polycrystalline data. Again the value of {$D_0$~=~9.2~(+3.4/-2.5)~$\times~10^{-5}\-$m$^2\-$s$^{-1}$} found in \cite{karunaratne2003interdiffusion} is in agreement with our results. 

\section{Conclusions}

We could show that aXPCS allows the observation of low concentrations of solute atoms if the electronic form-factor contrast between solvent and solute atoms is high enough. On the other hand it is the question of the coherent synchrotron brilliance to be able to measure still more dilute samples. Correlated diffusion according to the encounter model with correction for short-range order effects is a simple atomic diffusion process and the measured factors are reliable and in very good agreement with the results of tracer methods. Diffusion rates measured by aXPCS are, however, in the order of 10$^{-23}~ $m$^2$s$^{-1}$ and therefore on the lower border of what can be reached by the best tracer measurements. For measuring diffusivities this low, aXPCS is the only method capable of determining atomic jump vectors. 
 
The Pt-Pt repulsion indicated by the Monte Carlo simulation is in agreement with the Ni$_{3}$Pt phase \cite{cadeville1986magnetism} for low platinum concentration. Although for an atomic concentration as small as 3\% one can not speak of a Ni$_{3}$Pt phase, the characteristic atomic interactions which lead to its formation are present. It was, however, due to insufficient accuracy not yet possible to draw conclusions about solute-vacancy interaction from the comparison with simulations. With brighter synchrotron sources and better statistical significance coming along with them, it should be possible to investigate these features in the near future. Using Monte Carlo simulations on the bases of atomic interaction potentials and/or cluster expansion, aXPCS shows promise in being a valid tool to confirm and test ab initio calculated potentials. 

X-ray photon correlation spectroscopy allows one to draw conclusions about the dominating diffusion mechanism. To distinguish subtle differences e.g.~whether the encounter model or a model with slightly different jump probabilities takes place requires much better statistics than we achieved in our experiment.

The experimental data from the polycrystalline sample showed generally good agreement in activation energy and diffusion rates with the tracer method. It seems also possible to reveal the dominating diffusion mechanism (in this case nearest-neighbour jumps) with these data. In order to prepare polycrystalline samples allowing for exact measurements of diffusion rates, efforts have to be made to prepare samples with sufficiently small crystallites and to eliminate crystallite growth effects. Especially aXPCS experiments on powder samples seem to be particularly promising in this regard. 

Overall we conclude that aXPCS is a very reliable method for investigating atomic diffusion for all types of alloys and provides a reasonable reference for ab initio simulations. 
\appendix

\ack
This research was funded by the Austrian Science Fund (FWF) contract P-22402. 

We thank F. Zontone and the whole team of beamline ID10A at the ESRF in Grenoble, S. Puchegger from Faculty Center for Nano Structure Research University of Vienna for the EDX analysis, D. Geist and C. Rentenberger from Physics of Nanostructured Materials for TEM analysis, J. Akbarzadeh for her help with the SAXS and M. Rohrer for support with metallography. Furthermore our thanks go to F. Gr\"ostlinger for his assistance during beamtimes and to Martin Leitner for preliminary ab initio calculations of the interaction constants.
Also we thank H. Sassik from the Technical University of Vienna for providing the cold boat. 

\section*{References}
\bibliographystyle{unsrt.bst} 
\bibliography{literature}

\end{document}